\documentstyle[prl,aps,epsf,eqsecnum,floats,preprint]{revtex}
\begin{document}
%
\def\pr#1#2#3{ {\sl Phys. Rev.\/} {\bf#1}, #2 (#3)}
\def\prl#1#2#3{ {\sl Phys. Rev. Lett.\/} {\bf#1}, #2 (#3)}
\def\np#1#2#3{ {\sl Nucl. Phys.\/} {\bf B#1}, #2 (#3)}
\def\cmp#1#2#3{ {\sl Comm. Math. Phys.\/} {\bf#1}, #2 (#3)}
\def\pl#1#2#3{ {\sl Phys. Lett.\/} {\bf#1}, #2 (#3)}
\def\apj#1#2#3{ {\sl Ap. J.\/} {\bf#1}, #2 (#3)}
\def\aop#1#2#3{ {\sl Ann. Phy.\/} {\bf#1}, #2 (#3)}
\def\nc#1#2#3{ {\sl Nuo. Cim.\/} {\bf#1}, #2 (#3)}
\newcommand{\beq}{\begin{equation}}
\newcommand{\eeq}{\end{equation}}
\newcommand{\aprime}[1]{#1^\prime}
\newcommand{\daprime}[1]{#1^{\prime\prime}}
\newcommand{\aoy}{\frac{\alpha}{y}}
\newcommand{\poy}{\frac{\varphi}{y}}
\newcommand{\noy}{\frac{\nu}{y}}
\newcommand{\dap}{\delta a^\prime}
\newcommand{\dapp}{\delta a^{\prime\prime}}
\newcommand{\dnp}{\delta n^\prime}
\newcommand{\dnpp}{\delta n^{\prime\prime}}
\newcommand{\dpp}{\delta \phi^\prime}
\newcommand{\dbp}{\delta b^\prime}
\newcommand{\Vdb}{{Ve^{\bar b \phi_0}(\delta b+\bar b\delta\phi)}}
\newcommand{\ie}{{\it i.e.}}
\preprint{\vbox{\hbox{UCSD/PTH 00--13}}}
\title{Adding Matter to Poincare Invariant Branes}
\author{Benjam\'\i{}n Grinstein,\thanks{e-mail address: bgrinstein@ucsd.edu}
Detlef R. Nolte,\thanks{e-mail address: dnolte@ucsd.edu}
and Witold Skiba\thanks{e-mail address: wskiba@ucsd.edu}}
\address{Department of Physics, 
University of California at San Diego, La Jolla, CA 92093}
\date{April 2000; Revised June 2000}

\maketitle
\begin{abstract}
A solution to the cosmological constant problem has been proposed in
which our universe is a 3-brane in a 5-dimensional spacetime. With a
bulk scalar, the field equations admit a Poincare invariant brane
solution regardless of the value of the cosmological constant
(tension) on the brane. However, the solution does not include matter
in the brane. We  find new exact static solutions with
matter density and pressure in the brane. We study small perturbations about
these solutions. None seem consistent with
observational cosmology. As a byproduct we find a class of new
matterless static solutions and a non-static solution which, curiously,
requires the string value for the dilaton coupling.
\end{abstract}

\section{Introduction}
\label{intro}
The cosmological constant problem has evaded solution since its
inception\cite{weinberg1}. If only, it has become more severe: the
triumph of Quantum Field Theories as the correct description of the
fundamental interactions came at the cost of additive contributions to
the cosmological constant at big and disparate scales. It is natural
to expect that the cosmological constant gets contributions of order
$M_P^4$, where $M_P$ is the Planck scale, from short distance
gravitational dynamics; $M_W^4$, where $M_W$ is the mass of the
$W$-boson, from the phase transition associated with electroweak
symmetry breaking; $\Lambda_{\rm QCD}^4$, where $\Lambda_{\rm QCD}$ is
the scale of Quantum Chromo Dynamics, from the chiral symmetry breaking
phase transition; etc. A proper solution to the problem has to explain
how all such contributions are cancelled to absurdly high
precision. 

An intriguing solution has been proposed\cite{kss1,adks1} in which
spacetime is five dimensional, but the observable universe is
constrained to a four-dimensional hypersurface, a ``3-brane.'' The
authors exhibit solutions to the field equations which give a flat,
Poincare invariant, brane regardless of the value of the cosmological
constant. The geometry of space includes naked singularities which are
four dimensional hypersurfaces on either side of the brane on which
spacetime ends. The significance of and consistency of theories with
these singularities remains unclear\cite{singularities}. It has been
suggested\cite{flln1} that the singularities hide a fine tuning
equivalent to that required to set the cosmological constant to
zero. In Ref.~\cite{lowzee} the gravitational action is modified by
including a Gauss-Bonnet term, of second order in the curvature
tensor, and it is found that the singularity can be smoothed out but
only at the price of a fine tuning. It has also been shown\cite{kss2}
that in some cases the field equations admit solutions that correspond
to an Einstein-de Sitter universe on the brane. It has been suggested
that this type of models may be derived from string
theory\cite{deAlwis:2000pr}.

The solution is however incomplete. There is no matter in the toy
model of \cite{kss1,adks1}. It is necessary to incorporate matter if
the solution is to be relevant to cosmology. This is a non-trivial
issue: the standard paradigm, namely the big-bang cosmology based on a
Friedman-Robertson-Walker (FRW) metric, is observationally very
successful, so one should aim at reproducing, or at least approximating
this paradigm once matter is added.

We investigate inclusion of matter into these models.  
We discover new,  exact solutions to the brane
models with matter. They are static and therefore incompatible with
observational cosmology. Since the solutions are not unique, it is
possible that other solutions exist which appropriately describe the
expansion of the universe. Therefore we look for time dependent
solutions by linearizing the field equations about our new
solutions. As will be seen, generically the small perturbations
correspond to propagating modes or to non-expanding universes (static
solutions). Perturbations about  special, non-generic backgrounds
yield expanding universes that seem, however, inconsistent with
observational cosmology.

The paper is organized as follows. In section~\ref{prelim} we briefly
review the basic equations and the models of \cite{kss1,adks1}. In
section~\ref{sec:matter1} we present our new solutions with non-zero
matter on the brane. In
section~\ref{sec:bigMperts} we perform a small perturbation analysis
about these new solutions. We present our conclusions in
section~\ref{sec:conclusions}.

The cosmology of brane models has been investigated in a number of
papers. A general formulation was given in Ref.~\cite{bdl1}. The work
in Refs.~\cite{RScosmo,cgrt2} is concerned with the cosmology of brane
models of the Randall-Sundrum type\cite{Randall:1999ee}. In addition,
Randall-Sundrum models to which scalars are added have been of
interest\cite{bulkscalars}. A method to generate solutions to the non-linear
field equations in  classes of Randall-Sundrum models
with scalars was given in Ref.~\cite{DeWolfe:1999cp}. However, there
has been little, and only very recent, work on the
cosmology of automatically Poincare invariant branes\cite{Pcosmo}. 

\section{Preliminaries}
\label{prelim}
We denote the coordinates of spacetime by $x^A$, $A=0,\dots,4$, and
often use $t=x^0$ and $y=x^4$. The 3-brane is located at $y=0$. The
class of spherically symmetric metrics we study is parameterized by
three functions of $t$ and $y$ only\cite{bdl1}
\beq
\label{eq:bdlmetric}
ds^2=G_{AB}dx^Adx^B=n^2(t,y)dt^2-a^2(t,y)d\vec x^2 -b^2(t,y)dy^2.
\eeq 
Fixing $y=0$ we see that the metric gives a flat FRW cosmology on the
brane with scale factor $R(t')=a(t(t'),0)$ where $dt'=n(t,0)dt$. We
will denote by $g_{\mu\nu}$, with $\mu,\nu=0,\ldots,3$, the induced
metric on the brane.

The models of Refs.~\cite{kss1,adks1} include a scalar field
$\phi$. The action is
\beq
S=\int d^5x\, \sqrt{G}\left[-R+\frac43(\nabla\phi)^2-\Lambda e^{\bar a
\phi}\right]
+\int d^4x\,\sqrt{-g}\left[-Ve^{\bar b\phi}\right].
\eeq
We have adopted the notation of Ref.~\cite{kss1}, save for the
constants $\bar a$ and $\bar b$ which we have adorned with a bar to
distinguish them from the metric components $a^2(t,y)$ and
$b^2(t,y)$. $R$ denotes the Ricci scalar.

The peculiar normalization of the scalar field is adopted from string
theory: when $\phi$ is a string theory dilaton its couplings are
fixed. In particular, in this normalization, $\bar b = 2/3$ at lowest
order. We are not interested solely in this particular set of string
theory inspired parameters, so we keep the values unspecified. 

The constants $\Lambda$ and $V$ represent the cosmological constant in
the bulk (5-dimensional space) and on the brane, respectively. For
our analysis we set $\Lambda=0$. As seen in Ref.~\cite{kss1}
this simplifies the analysis without compromising the essential
features of the model. Moreover, one could imagine that if the model
is embedded in a supersymmetric setting, $\Lambda$  could naturally
vanish. The cosmological constant problem is associated with standard
model fields' contributions to $V$, $V\sim M_P^4 +M_W^4 +\Lambda_{\rm
QCD}^4+\cdots$

Einstein's equations are
\beq
R^{AB}-\frac12G^{AB}R=\kappa^2T^{AB}.
\eeq
Here $R^{AB}$ and $R$ are the Ricci tensor and scalar. The
gravitational constant is $\kappa^2$ and from now on we work in units
of $\kappa^2=1$. $T^{AB}$ is the stress-energy tensor, which has two
components:
\beq
T^{AB}=\tilde T^{AB}+\frac{S^{AB}}{b}\delta(y),
\eeq
where $\tilde T^{AB}$ is derived as usual by varying the action with
respect to the metric, and $S^{AB}$ is a contribution from a perfect
fluid of density $\rho$ and pressure $p$ on the brane,
\beq
S^A_B=\mbox{diag}\,(\rho,-p,-p,-p,0).
\eeq
Alternatively one may take the matter fluid to couple to the scalar,
thus
\beq
S^A_B=\mbox{diag}\,(e^{\bar b \phi}\rho,
 -e^{\bar b \phi}p,-e^{\bar b \phi}p,-e^{\bar b \phi}p,0).
\eeq
These two ways of writing $S^A_B$ 
correspond to distinct physical models. However the
distinction turns out to be irrelevant for the exact solutions that we
present in Sec.~\ref{sec:matter1}.
The field equation for the scalar $\phi$ is
\beq
-\frac83\nabla^2\phi-
\frac{\sqrt{-g}}{\sqrt{G}}\delta(y)\bar bVe^{\bar b\phi}=0.
\eeq

For the particular metric (\ref{eq:bdlmetric}) Einstein's equations and the
$\phi$ field equation are
\begin{eqnarray}
3\,\left[ \,\left( 
            (\frac{\dot{a}}{a})^2 + 
            \frac{\dot{a}\,\dot{b}}{a \, b} \right) 
         +  \frac{n^{2}}{b^{2}} \,\left(  -
            \frac{\daprime{a}}{a}        -
            (\frac{\aprime{a}}{a})^2  +
            \frac{\aprime{a}\,\aprime{b}}{a \, b}
            \right) \right] & = & 
 \frac{2}{3} n^2 \left( 
          \frac{\dot{\phi}^2}{n^2} + \frac{{\aprime{\phi}}^2}{b^2} \right)  + 
          \delta(y) \,\, \frac{n^2}{b} (\frac{1}{2} V e^{\bar b \phi} + \rho),
          \label{Einstein00} \\
3\,\left( 
            \frac{\dot{a}\,\aprime{n}}{a \, n} + 
            \frac{\aprime{a}\,\dot{b}}{a \, b} - 
            \frac{\aprime{\dot{a}}}{a}   \right) & = &
\frac{4}{3} \dot{\phi} \aprime{\phi},  \\
\frac{a^{2}}{n^{2}}\,\left( -
             \frac{\dot{a}^{2}}{a^{2}} -2 
             \frac{\ddot{a}}{a} + 2
             \frac{\dot{a}\,\dot{n}}{a \, n} - 2
             \frac{\dot{a}\,\dot{b}}{a \, b} -
             \frac{\ddot{b}}{b} +
             \frac{\dot{n}\,\dot{b}}{n \, b} \right) + & & \nonumber \\
             \frac{a^{2}}{b^{2}}\,\left( 
             (\frac{\aprime{a}}{a})^2 +2
             \frac{\daprime{a}}{a} + 2
             \frac{\aprime{a}\,\aprime{n}}{a \, n} - 2
             \frac{\aprime{a}\,\aprime{b}}{a \, b} +
             \frac{\daprime{n}}{n} -
             \frac{\aprime{n}\,\aprime{b}}{n \, b}  \right)  & = &   
\frac{2}{3} a^2 \left( 
        \frac{\dot{\phi}^2}{n^2} - \frac{{\aprime{\phi}}^2}{b^2} \right)  - 
        \delta(y) \,\, \frac{a^2}{b} ( \frac{1}{2} V e^{\bar b \phi} - p),
  \label{Einstein11} \\
3 \,\left[ %
            \frac{b^{2}}{n^{2}} \,
        \left(  -
             \frac{\dot{a}^{2}}{a^{2}} - 
             \frac{\ddot{a}}{a} + 
             \frac{\dot{a}\,\dot{n}}{a \, n}
         \right) +
          \,\left(
             \left(\frac{\aprime{a}}{a}\right)^2 + 
              \frac{\aprime{a}\,\aprime{n}}{a \, n} 
          \right) 
     \right] & = &
\frac{2}{3} b^2 \left( 
          \frac{\dot{\phi}^2}{n^2} + \frac{{\aprime{\phi}}^2}{b^2} \right), 
  \label{Einstein44} \\
 \frac{8}{3} \frac{1}{b^2}
 \left[ \left( \ddot \phi \frac{b^2}{n^2} - \daprime \phi \right)
       -\aprime \phi \left( 3 \frac{\aprime a}{a} - \frac{\aprime b}{b} +
                      \frac{\aprime n}{n} \right) \right.
       + & & \nonumber \\
  \left.
  \dot \phi \frac{b^2}{n^2} \left(3 \frac{\dot a}{a} + \frac{\dot b}{b}
                                   - \frac{\dot n}{n} \right) \right]
  &=& - \delta(y) \,\, \frac{1}{b}  
    V \, \bar{b} \, e^{\bar{b} \phi} .
  \label{phieq0}
\end{eqnarray}            
Here a dot is a shorthand for $\partial/\partial t$ and a prime for
$\partial/\partial y$. The first four equations correspond to the 00,
04, 11 and 44 components of Einstein's equations.

Conservation of the stress-energy tensor would be automatic were it
derived from a local action. However, since a fluid component has been
added on the brane, the equation $T^{AB}_{\hspace{5mm} ;B}=0$
contains additional non-trivial information.
Conservation of energy, $T^{0B}_{\hspace{4mm} ;B}=0$, gives
\begin{eqnarray}
  \frac{8}{3} \frac{1}{b^2} \dot\phi \left[ 
       \left( \ddot \phi \frac{b^2}{n^2} - \daprime \phi \right) - 
       \aprime \phi \left( 3 \frac{\aprime a}{a} - \frac{\aprime b}{b} +
                      \frac{\aprime n}{n} \right) + \right. & & \\
  \left. \dot \phi \frac{b^2}{n^2} \left(3 \frac{\dot a}{a} + \frac{\dot b}{b} 
                                   - \frac{\dot n}{n} \right)
  \right] & = & 
   -\frac{\delta(y)}{b}\left[ \dot \phi 
      V \, \bar{b} \, e^{\bar{b} \phi}+  
  2 \left( \dot \rho + 
      3\frac{\dot a}{a} (\rho +p ) \right)\right] \nonumber 
\end{eqnarray}
while $T^{4B}_{\hspace{4mm} ;B}=0$ yields
\begin{eqnarray}
\label{eq:T4AAexplicit}
 \frac{8}{3} \frac{1}{b^2} \aprime\phi\left[ 
  \left( \ddot \phi \frac{b^2}{n^2} - \daprime \phi \right)
       -\aprime \phi \left( 3 \frac{\aprime a}{a} - \frac{\aprime b}{b} +
                      \frac{\aprime n}{n} \right) + \right. & & \\
   \left. \dot \phi \frac{b^2}{n^2} 
      \left(3 \frac{\dot a}{a} + \frac{\dot b}{b} 
     - \frac{\dot n}{n} \right) \right] 
  & =  &  \delta(y) \, \, \frac{1}{b} \left( \frac{\aprime n}{n} 
        (V e^{\bar{b} \phi}+2 \rho)
         + 3\frac{\aprime a}{a} (V e^{\bar{b} \phi} -2 p) \right). \nonumber
\end{eqnarray}
Using the field equation for the scalar, Eq.~(\ref{phieq0}), in the
conservation of energy equation gives
\beq
\dot\rho+3\frac{\dot a}a(\rho+p)=0
\eeq
on the brane.
The second conservation equation, $T^{4B}_{\hspace{4mm} ;B}=0$,
is always satisfied as a consequence of the $\phi$ equation of motion
and Einstein's equations. Only the brane part of
$T^{4B}_{\hspace{4mm} ;B}=0$ does not follow immediately from
Eq.~(\ref{phieq0}). The identity for $y=0$ can be confirmed by
taking the $y$ derivative of Eq.~(\ref{Einstein44})
and then using Eqs.~(\ref{Einstein00}) and (\ref{Einstein11}).

It is stated in Ref.~\cite{bdl1} that for brane geometries of the form
given by Eq.~(\ref{eq:bdlmetric}) the equation of conservation of
transverse momentum, $T^{4A}{}_{;A}=0$,  reduces, on the brane, to 
\beq
\label{eq:fakeT4AA}
\frac{n'}n\rho=3\frac{a'}ap.  
\eeq 
It is understood here that when discontinuous quantities are
evaluated on the brane, like $n'$ and $a'$, they are given by
their average, \ie, 
\beq
n'(y=0)=\frac12(n'(y=0+)+n'(y=0-)).
\eeq
The equation is seldom considered any further because, for $Z_2$
symmetric brane-spaces, that is for  metrics with $y\to-y$ symmetry,
which are overwhelmingly most common,  the averages both vanish
separately, $a'=0=n'$, and the equation is trivially satisfied. 

However the solutions we consider are not $Z_2$ symmetric. This is
also true of the solutions in Refs.~\cite{kss1} but there
Eq.~(\ref{eq:fakeT4AA}) is still trivially satisfied since the matter
density and pressure both vanish. It is easy to see that our solutions
below, Eqs.~(\ref{eq:bigsolsa})--(\ref{eq:bigsolsphi}), do not satisfy
Eq.~(\ref{eq:fakeT4AA}). The reason is, in fact, that
Eq.~(\ref{eq:fakeT4AA}) does not apply to the case in which there are
bulk scalars. The conservation of $y$-momentum is automatically
satisfied in the bulk because the action is translation invariant in
the bulk. So at issue here is only the conservation equation on the
brane. Retaining only the bulk terms involving second derivatives, the
conservation Eq.~(\ref{eq:T4AAexplicit}) gives
\beq
\label{eq:T4AAcorrect}
-\frac43\phi'\phi'' + b\delta(y) \left[-\frac{n'}{n}\rho +
3\frac{a'}ap - 
\frac12Ve^{\bar b \phi}\left(\frac{n'}{n}+3\frac{a'}a\right)\right]=0.
\eeq
Following Ref.~\cite{bdl1} we interpret the discontinuous derivatives
on the  brane as averages, and this will remain implicit in what
follows. Integrating gives a jump equation,
\beq
-\frac2{3b}\Delta\left(\phi'\right)^2=
\frac{n'}{n}\rho - 3\frac{a'}ap + 
\frac12Ve^{\bar b \phi}\left(\frac{n'}{n}+ 3\frac{a'}a\right)
\eeq
We have verified that our solutions,
Eqs.~(\ref{eq:bigsolsa})--(\ref{eq:bigsolsphi}), and also the
perturbations, Eqs.~(\ref{eq:solbulkphi})--(\ref{eq:solbulkn})
and~(\ref{eq:jumpeqspertgrala})--(\ref{eq:jumpeqspertgralphi}) satisfy
this equation. In fact one can prove this without reference to
the explicit form of the solution. 

In Ref.~\cite{sundrum1} it is advocated that the correct form of the
conservation of transverse momentum equation is 
\beq
\label{eq:sundrum}
T^{4\mu}{}_{,\mu}=0,
\eeq
where the index $\mu$ runs from $0$ to $3$ only. However, the first
term in Eq.~(\ref{eq:T4AAcorrect}), involving the all important second
derivative term $\phi''\phi'$, arises from the derivative
$T^{44}{}_{,4}$, which is ommitted from Eq.~(\ref{eq:sundrum}).

Let us now describe the models studied in Refs.~\cite{kss1,adks1}.
They take $\rho=p=0$. The solutions all have $n(t,y)=a(t,y)$ and
$b(t,y)=1$, and are static, $\dot a=\dot\phi=0$. For example, case I
studied in Ref.~\cite{kss1} has $\Lambda=0$ and solutions
\begin{eqnarray}
n=a & = & \cases{(1-y/y_+)^{\gamma_+}, & for $y>0$\cr
           (1-y/y_-)^{\gamma_-}, & for $y<0$\cr}\\
\phi & = & \cases{\varphi_+\log(1-y/y_+)+c, & $y>0$\cr
                  \varphi_-\log(1-y/y_-)+c, & $y<0$\cr}.
\end{eqnarray}
The constants $\gamma_\pm=1/4$ and $|\varphi_\pm|=3/4$ are fixed by
the field equations in the bulk. Case I has, in particular,
$\varphi_\pm=\mp 3/4$. The constants $y_\pm$ are determined by
``jump'' conditions, that is, by requiring that the second derivatives
of fields in the field equations correctly reproduce the
$\delta$-function terms from the brane. The constant $c$ is an
irrelevant constant shift of the scalar field.


\section{Solutions with Matter}
\label{sec:matter1}
The models of Refs.~\cite{kss1,adks1} provide a solution to the
cosmological constant problem which is deficient in several ways:
(1)~There are naked singularities. Whether these are problematic
remains an open question; see, for example,
Refs.~\cite{singularities,flln1}. (2)~There is a massless scalar which
interacts with all matter with a universal, gravity-like coupling
strength. This is ruled out \cite{gravityexps} unless the coupling is made
sufficiently weak. It can be arranged, however, by choosing the
parameter $\bar b$ small enough. (3)~It describes a static cosmology,
in conflict with observation (see, however,
Ref.~\cite{burbridge}). (4)~It does not include matter density (and
pressure) on the brane.

Here we address the last two of these problems. One hopes the two are
connected: when matter is included in the model the universe will
evolve in time. Of course, not only should the universe evolve, but
the rate of expansion should be adequate.

However, even after the introduction of matter the model admits static
solutions. This is somewhat surprising, particularly if it is
contrasted with the FRW cosmology which admits a static solution only
if the matter density is precisely balanced by a cosmological term
giving the Einstein universe. 
Since we are ultimately interested in time dependent solutions, we
will look at time dependent small fluctuations about these solutions
in the next section.

We look for solutions to the field equations in the bulk with the
ansatz
\begin{eqnarray}
\label{eq:ansatz}
a & = & y^\alpha,\\
n & = & y^\nu,\\
b & = & 1,\\
\phi & = & \varphi\log y.
\end{eqnarray}
The $\phi$ field equation gives
\beq
\label{eq:nugiven}
3\alpha+\nu=1.
\eeq
All Einstein field equations give then 
\beq
2\alpha^2-\alpha+\frac29\varphi^2=0.
\eeq
For definiteness we consider solutions akin to case I of
Ref.~\cite{kss1}, with
\beq
\label{eq:varphigiven}
\varphi=\mp3\sqrt{\alpha/2-\alpha^2},
\eeq
where the upper and lower signs correspond to the regions $y>0$ and
$y<0$, respectively.

The full solution is found by shifting $y$ by $y_+$($y_-$) on
$y>0$($y<0$), and pasting these using the jump equations. We have
\begin{eqnarray}
\label{eq:bigsolsa}
a & = & A\left(1-\frac{y}{y_\pm}\right)^{\alpha_\pm},\\
\label{eq:bigsolsn}
n & = & N\left(1-\frac{y}{y_\pm}\right)^{\nu_\pm},\\
\label{eq:bigsolsphi}
\phi & = & \varphi_\pm \log(1-y/y_\pm)+c.
\end{eqnarray}
Here $A$ and $N$ are arbitrary constants that can be set to unity by a
coordinate rescaling.
The jump equations are
\begin{eqnarray}
\label{eq:jumpeqsgral}
\frac{\Delta a'}{a} & = &-\frac{b}3\left(\frac12Ve^{\bar b \phi}+\rho\right),\\
2\frac{\Delta a'}{a}+\frac{\Delta n'}{n} & = &
b\left(-\frac12Ve^{\bar b \phi}+p\right), \\
\Delta\phi' & = &\frac38b\bar bVe^{\bar b \phi},
\end{eqnarray}
where $\Delta a'=a'(y=0+)-a'(y=0-)$, etc. These give three equations
for five unknowns,
\begin{eqnarray}
\label{eq:jumpone}
\frac{\alpha_+}{y_+}-\frac{\alpha_-}{y_-} & = &
\frac13\left(\frac12Ve^{\bar b c}+\rho\right),\\
\label{eq:jumptwo}
\frac{1}{y_+}-\frac{1}{y_-} & = &-p+\frac13\rho+\frac23Ve^{\bar b
c},\\
\label{eq:jumpthree}
\frac{\sqrt{\alpha_+/2-\alpha_+^2}}{y_+}+\frac{\sqrt{\alpha_-/2-\alpha_-^2}}{y_-}
& = &\frac18\bar b Ve^{\bar b c}.
\end{eqnarray}
These can always be solved for three unknowns (say $\alpha_+$ and
$y_\pm$) in terms of $\rho$, $p$ and two other unknowns (say
$\alpha_-$ and $c$). The reason not all unknowns are determined is
twofold. First, gauge (diffeomorphism) invariance allows us to make
unphysical changes to our solutions. This will be explained below in
detail, but for now it suffices to know that  one may fix the gauge
freedom by setting, say, $c=0$. And secondly, even for vanishing
$\rho$ and $p$ one can find a class of solutions parameterized by one
parameter. These are new solutions to the model considered in
\cite{kss1}, which could not be discovered there because it was
assumed that $n=a$. Indeed, imposing this one has $\nu=\alpha$ which
together with Eqs.~(\ref{eq:nugiven}) and~(\ref{eq:varphigiven}) imply
\beq
\alpha=\nu=\frac14
\eeq
and
\beq
\varphi=\mp\frac34
\eeq
as found in case I of \cite{kss1}.

We have searched for time dependent solutions. We found one special
solution, valid only for $\rho=0$ and provided $\bar b=2/3$.  It is
given by
\begin{eqnarray}
\label{eq:grotesque1}
ds^2&= & (1-y/y_\pm)^2dt^2-(1-y/y_\pm)^2t^2d\vec x^2-t^2dy^2,\\
\label{eq:grotesque2}
\phi&= & -\frac32\ln[t(1-y/y_\pm)]+c,\\
\frac1{y_+}-\frac1{y_-}& = & \frac16Ve^{2c/3}.
\end{eqnarray}
Perplexingly, the special value $\bar b=2/3$
corresponds to the string theory value of this parameter. Other
non-static solutions were given in Ref.~\cite{kss2}. In this paper we
will not consider these type of non-static matter-free solutions
further.

\section{Small perturbations about Large Matter Density}
\label{sec:bigMperts}
Armed with the new solutions with static matter density, we proceed to
investigate the time dependence of small matter perturbations. Let us
denote the static solution of the previous section by $n_0$, $a_0$,
$b_0$, $\phi_0$, $\rho_0$ and $p_0$. We look for solutions to the
field equations,
Eqs.~(\ref{Einstein00})--(\ref{phieq0}), of the form
\begin{eqnarray}
\label{eq:perturbs}
n & = & n_0 (1+\delta n),\nonumber\\
a & = & a_0 (1+\delta a),\nonumber\\
b & = & b_0 (1+\delta b),\\
\phi & = & \phi_0+\delta \phi,\nonumber\\
\rho & = & \rho_0 + \delta\rho,\nonumber\\
p & = & p_0 + \delta p.\nonumber
\end{eqnarray}

We count orders of the perturbative expansion parametrically in
$\delta \rho$ and $\delta p$. That is, we
re-scale $\delta \rho\to\epsilon \delta \rho$, count powers of
$\epsilon$ and set $\epsilon=1$ at the end of the calculation.
In particular this implies
that we make no assumption as to the relative importance of temporal
or spatial derivatives\cite{cgrt2}.

To derive the linearized equations in the bulk, we use again this
parameterization and  the
explicit form of the zeroth order solutions,
Eq.~(\ref{eq:ansatz}). The 00, 04, 11 and 44 components of Einstein's
equations and the $\phi$ field equation give
\begin{eqnarray}
  \dapp+4\aoy\dap-\aoy\dbp+\frac49\poy\dpp& = & 0,\\
  \frac\partial{\partial t}\left(\noy\delta a+\aoy\delta b-\aoy\delta a
        -\dap-\frac49\poy\delta\phi\right) & = & 0,\\
  2\dapp+\frac2y\dap+2\frac{\alpha+\nu}y\dnp+\dnpp
     -\frac{2\alpha+\nu}y\dbp+\frac43\poy\dpp
     & = & \frac1{n_0^2}\left(2\delta\ddot a+\delta \ddot b\right), \\
  \frac{2\alpha+\nu}y\dap+\aoy\dnp-\frac49\poy\dpp
     & = & \frac1{n_0^2}\delta\ddot a,\\
  \frac1y\frac{\partial}{\partial y}\left[y\dpp+\varphi(3\delta a+
     \delta n-\delta b)\right] & = & \frac1{n_0^2}\delta\ddot \phi.
\end{eqnarray}

The solution to these equations gives $\delta \phi$, $\delta b$ and
$\dnp$ in terms of $\delta a$:
\begin{eqnarray}
\label{eq:solbulkphi}
\delta \phi & = & \frac1\alpha(\varphi\delta a -F),\\
\label{eq:solbulkb}
\delta b & = &
\frac1\alpha\left[(y\delta a)^\prime-\frac49\frac\varphi{\alpha} F
-\xi\right],\\
\label{eq:solbulkn}
\dnp & = &
\frac1\alpha\left[\frac{y}{n_0^2}\delta \ddot a-\frac{\partial}{\partial y}
\left((3\alpha-1)\delta a +
\frac49\frac\varphi\alpha F\right)\right].
\end{eqnarray}
Here $\xi$ is an arbitrary  constant and $F$ is a function of $t$ and $y$
satisfying
\beq
\label{eq:waveequation}
\frac1y\frac{\partial}{\partial y}\left(y \frac{\partial F}{\partial
y} \right ) - \frac1{n_0^2}\frac{\partial^2 F}{\partial t^2}=0.
\eeq

We connect the bulk solutions for $y>0$ and $y<0$ demanding continuity
of the fields at the brane, $y=0$, and using the jump equations
(\ref{eq:jumpeqsgral}) for the discontinuous derivatives at $y=0$. The
latter give jump conditions for the perturbations:
\begin{eqnarray}
\label{eq:jumpeqspertgrala}
\Delta\dap &=&-\frac16\Vdb-\frac13\delta\rho-\frac13\rho_0\delta b, \\
\label{eq:jumpeqspertgraln}
\Delta\dnp &= &-\frac16\Vdb
         +\delta p+\frac23\delta\rho+(p_0+\frac23\rho_0)\delta b,  \\
\label{eq:jumpeqspertgralphi}
\Delta\dpp &=&\frac38\bar b\Vdb.
\end{eqnarray}
It must be observed that both $\xi$ and $F$ can have different values
on either side of the brane and, moreover, that $F$ may be
discontinuous at $y=0$. In addition, conservation of energy gives, on
the brane,
\beq
\label{eq:energyconperts}
\delta\rho+3(\rho_0+p_0)\delta a=\delta\rho_0, 
\eeq 
where $\delta\rho_0$ is a constant of integration.

There is a gauge freedom, that is, reparameterization invariance
consistent with the form of our metric. Starting from the metric
\beq
ds^2=n^2(t',y')dt'^2-a^2(t',y')d\vec x^2 -b^2(t',y')dy'^2,
\eeq
we look for infinitesimal transformations
\begin{eqnarray}
t'&=&t+T(t,y)\\
y'&=&y+Y(t,y)
\end{eqnarray}
that leave the form of the metric invariant. Here $T$ and $Y$ are
infinitesimal. The only constraint on these functions comes from the
absence of off-diagonal terms in the metric:
\beq
\label{eq:YTcondition}
n^2T'-b^2 \dot Y=0.
\eeq
Under the gauge transformation the metric variations are
\begin{eqnarray}  
\label{eq:gaugeTs}
\delta n & = & \frac{n'_0}{n_0}Y+\dot T ,\\
\delta a & = & \frac{a'_0}{a_0}Y, \\
\delta b & = & Y', \\
\delta\phi & = & \phi'_0 Y .
\label{eq:gaugephi}
\end{eqnarray}
For simplicity we have indicated the variation about a static solution
with $b_0=1$ and $\dot a_0=\dot n_0 =\dot\phi_0=0$. It is instructive
to check that our solutions of the field equations for the
perturbations are invariant under these transformations, that is, that
the perturbations (\ref{eq:gaugeTs})-(\ref{eq:gaugephi})
satisfy the field equations
automatically. Indeed, the solution for $\delta \phi$,
Eq.~(\ref{eq:solbulkphi}), is satisfied provided one takes
$F=0$. Then, the solution for $\delta b$, Eq.~(\ref{eq:solbulkb}),
requires $\xi=0$. Finally the solution for $\dnp$,
Eq.~(\ref{eq:solbulkn}), is satisfied provided
\[
\dot T' -\frac1{n_0^2}\ddot Y = 0 ,
\]
which is a consequence of the condition~(\ref{eq:YTcondition}).

One may fix the gauge by imposing, for example,
\beq
\label{eq:gauegfix}
\delta b(y,t)=0\qquad\mbox{and}\qquad \delta\phi(y=0,t)=0.
\eeq
There is some residual gauge freedom: there are further
transformations with $Y=0$ and $T=T(t)$. These are
uninteresting time reparameterizations. Were we to impose a gauge
condition on any of the fields that have a discontinuous derivative on
the brane, we could only require that it vansihes for either $y>0$ or
$y<0$, but not both, since the reparametrization functions are
smooth. 

We are ready to present our solution to the linearized field
equations. The solutions to the bulk equations express $\delta\phi$,
$\delta b$ and $\delta \aprime{n}$ in terms of $\delta a$ and
$F$. Once $F$ is determined in terms of $\delta a$ our task is to
determine $\delta a$ only. We will proceed as follows. First we use
the gauge conditions and our solutions in the bulk,
Eqs.~(\ref{eq:solbulkphi})--(\ref{eq:solbulkn}), to eliminate the
function $F$ everywhere, and to relate $\delta \aprime{a}$ to $\delta
a$ on the brane. Then we consider the jump equations which impose
further restrictions on $\delta a$ and its derivatives on the
brane. These constraints on the brane are used, finally, to determine
$\delta a$ in the bulk. The cosmlogy depends only on the fields on the
brane. Therefore, we concentrate on determining as fully as possible
the fields on the brane.

We eliminate the function $F$ by fixing the gauge as in
Eq.~(\ref{eq:gauegfix}) and using our solution of the field equations
in the bulk, Eq.~(\ref{eq:solbulkb}), thus
\beq
\label{eq:Ffix1}
F=\frac94\frac\alpha\varphi \left[(y\delta a)'-\xi\right].
\eeq
We can now fix $\delta \aprime{a}$ on the brane in terms of  $\delta a$.
Consider the constraints from requiring
continuity of $\delta \phi$ at $y=0$. We
have in addition the gauge choice in
Eq.~(\ref{eq:gauegfix}), $\delta\phi(y=0,t)=0$, so we obtain
\begin{eqnarray}
\label{eq:appgiven}
(4\alpha_+-1)\delta a &=&  (y_+\delta\aprime{a}_++\xi_+),\\
\label{eq:apmgiven}
(4\alpha_--1)\delta a &=&  (y_-\delta\aprime{a}_-+\xi_-).
\end{eqnarray}
Here and below we denote the limiting values of fields as $y\to0\pm$
with a corresponding subscript, eg, $\delta\aprime{a}_\pm\equiv
\lim_{y\to0\pm}\delta\aprime{a}$. 

The first jump equation, Eq.~(\ref{eq:jumpeqspertgrala}), gives a
constraint between the constants of integration. From the jump
equations for the exact solution, Eqs.~(\ref{eq:jumpone})
and~(\ref{eq:jumptwo}), one has
\beq
\label{eq:condition}
\rho_0+p_0=\frac{4\alpha_+-1}{y_+}+\frac{4\alpha_--1}{y_-}.  
\eeq 
Using this and Eqs.~(\ref{eq:appgiven}) and~(\ref{eq:apmgiven}) 
in Eq.~(\ref{eq:jumpeqspertgrala}) we obtain
\beq
\frac13\delta\rho_0=\frac{\xi_+}{y_+}+\frac{\xi_-}{y_-}.
\eeq
The remaining two jump equations, (\ref{eq:jumpeqspertgraln})
and~(\ref{eq:jumpeqspertgralphi}), involve $\delta\daprime{a}_\pm$ in
addition to $\delta\aprime{a}_\pm$, $\delta{\ddot a}$
and~$\delta{a}$. They can be solved to give, on the brane,
$\delta\daprime{a}_\pm$ in terms of $\delta{\ddot a}$ and
$\delta{a}$. However, the resulting expresions are  long and not
terribly illuminating so we refrain form 
presenting them here.

The jump equations do not fix the time behavior of $\delta a$ on the
brane. We now describe a procedure that determines the time behavior
and the bulk dependence of $\delta a$. Note that since $F$ is given in
terms of $\delta a$ through Eq.~(\ref{eq:Ffix1}), one must now impose
that $\delta a$ satisfy the wave-like
equation~(\ref{eq:waveequation}). The solution to
Eq.~(\ref{eq:waveequation}) is straightforward,
\beq
\label{eq:fourieryt}
F_\pm(y,t)=\int d\omega\, e^{i\omega t}\tilde F_\pm(\omega) 
\frac{J_0(\frac{\omega}{3\alpha_\pm}|y-y_\pm|^{3\alpha_\pm})}%
{J_0(\frac{\omega}{3\alpha_\pm}|y_\pm|^{3\alpha_\pm})},
\eeq
where $J_0$ is a Bessel function. There is a second solution involving
the Neuman function. However, we have dismissed it since it is arbitrarily
large as $|y-y_\pm|\to0$, which is outside the validity of
perturbation theory. The Fourier coefficient $\tilde F_\pm(\omega)$ is
determined by the function on the brane, $F(0,t)$. Using the
conditions on derivatives of $\delta a$ on the brane one may express
this solely in terms of $\delta a$:
\beq
\frac92\frac{\alpha_\pm(1-2\alpha_\pm)}{\varphi_\pm}\delta a(0,t)
=\int d\omega\, e^{i\omega t}\tilde F_\pm(\omega) .
\eeq

Thus, knowledge of $\delta a$ on the brane determines all fields
everywhere. The time-dependence of $\delta a$, however, is not
arbitrary. Consistency of Eq.~(\ref{eq:fourieryt}) and the jump
equations gives an integro-differential equation for $\delta a$ on the
brane. To see this, differentiate Eq.~(\ref{eq:fourieryt}) and
evaluate at $y=0$:
\beq
-y_\pm\delta\daprime{a}_\pm+2\delta\aprime{a}_\pm =
\int d\omega\, e^{i\omega t} (4\alpha_\pm-2)\delta \tilde a(\omega)
\omega |y_\pm|^{3\alpha_\pm-1}
\frac{J_1(\frac{\omega}{3\alpha_\pm}|y_\pm|^{3\alpha_\pm})}%
{J_0(\frac{\omega}{3\alpha_\pm}|y_\pm|^{3\alpha_\pm})},
\eeq
where $(1-2\alpha_\pm)\delta \tilde a(\omega)=2\varphi_\pm/9\alpha_\pm
\tilde F_\pm(\omega)$.
Now the left hand side of this equation can be expressed in terms of
$\delta\ddot a$ and $\delta a$ through the jump equations. One obtains
\beq
-y_\pm\delta\daprime{a}_\pm+2\delta\aprime{a}_\pm = A_\pm \delta\ddot
a +B_\pm \delta a + C_\pm,
\eeq
where $A_\pm$ and $B_\pm$ are constants given in terms of the
parameters of the background solution, and $C_\pm$ also depends on
these parameters but is, in addition, linear in the small constants
$\xi_\pm$. We have ommitted the complicated and long expressions for
these. Thus, we obtain an equation that determines the time dependence
of $\delta a$
\beq
A_\pm \delta\ddot a +B_\pm \delta a + C_\pm =
\int d\omega\, e^{i\omega t} (4\alpha_\pm-2)\delta \tilde a(\omega)
\omega |y_\pm|^{3\alpha_\pm-1}
\frac{J_1(\frac{\omega}{3\alpha_\pm}|y_\pm|^{3\alpha_\pm})}%
{J_0(\frac{\omega}{3\alpha_\pm}|y_\pm|^{3\alpha_\pm})}.
\eeq

A particular solution to this eqution is easily found by Fourier
transform:
\beq
\delta a(0,t) = -C_\pm/B_\pm.
\eeq
To this solution one may add arbitrary linear combinations of
solutions to the associated homogeneous equation (obtained by setting
$C_\pm=0$). These solutions have time dependence of the form of simple
exponentials, $\exp{i\omega_0 t}$, with characterisitic frequencies $\omega_0$
that are solutions to
\beq
-\omega_0^2 A_\pm +B_\pm-(2-4\alpha_\pm)
\omega_0 |y_\pm|^{3\alpha_\pm-1}
\frac{J_1(\frac{\omega_0}{3\alpha_\pm}|y_\pm|^{3\alpha_\pm})}%
{J_0(\frac{\omega_0}{3\alpha_\pm}|y_\pm|^{3\alpha_\pm})}=0.
\eeq
There is an infinite number of solutions to this equation. The
function $J_0$ has an infinite number of simple zeros. Between two
succesive zeros the last term takes on any value. There may be
additional complex solutions when $A_\pm B_\pm<0$. 

The physical interpretation of our solution is straightforward. The
infinite set of oscilatory modes simply correspond to field
excitations in the bulk. Neither their amplitudes nor their frequencies
depend on the additional density perturbation $\delta\rho_0$. 
This is to be expected because even in the absence of additional
matter there can be propagating gravitational and scalar field
waves. It is only the
particular 
solution that actually depends on the added matter. Ignoring the possible
excitation of propagating modes, we see that the new solution is
static.

There are some caveats. This conclusion does not hold if the parameters
of the background are such that $B_\pm=0$ or if there are complex
solutions $\omega_0$. In the former case, if in addition $C_\pm=0$,
there are further
solutions of the form
\beq 
\delta a \propto t.
\eeq
This is of particular interest beacuse the perturbations may now be
time dependent. In fact, such time dependent perturbations are
expected since they must correspond to the linearization of the exact
solutions of Ref.~\cite{kss2} for which
$a\propto\exp(f(y)+\sqrt{\bar\Lambda}t)\simeq1+f(y)+\sqrt{\bar\Lambda}t$,
where $\bar\Lambda$ is an arbitrary constant. We have verified that
$B_\pm=0$ and $A_\pm\ne0$ for parameters corresponding to the
background of Ref.~\cite{kss2} ($\alpha_\pm=1/4$,
$\varphi_\pm=\mp3/4$, $\rho_0=\delta \rho_0=0$).  
There are also solutions of the form
$\delta a \propto t^2$. If there are modes of complex $\omega_0$, the
solutions are not simply oscilatory, but will in addition have
exponential dependence. None of these solutions are suitable for
observational cosmology.

\section{Conclusions}
\label{sec:conclusions}
The proposed solution to the cosmological constant problem of
Refs.~\cite{kss1,adks1} is incomplete in that it does not include
matter on the brane, \ie, in our universe. 
We have found new exact solutions including matter on the
brane. These solutions are static and therefore describe an
unacceptable cosmology.

However, the equations admit other solutions even under the same
assumptions on the symmetry of the metric. This had been recognized in
Ref.~\cite{kss2} which found curved brane solutions to the model
dubbed case I in Ref.~\cite{kss1}. Here we have studied new solutions
obtained as small perturbations about our new static solutions with
matter. The small perturbations that are not simply propagating waves
are generally static. In special cases the non-propagating,  
small perturbations grow linearly with time, but we have identified
these as the De-Sitter-like solutions of Ref.~\cite{kss2}.
On this basis it is tempting to rule out these as viable
cosmologies. 

In standard FRW cosmology the evolution of the scale factor is
completely determined once the equation of state is fixed. Once
spherical symmetry is chosen and matter is specified, Einstein's
equations determine the cosmology. However, this is not the case in the
peculiar brane cosmologies of Refs.~\cite{kss1,adks1}. To understand
what is happening one could consider this as an initial value
problem. In standard cosmology if the metric and matter content were
specified at an initial time (in a fixed gauge), one could evolve
forward using the field equations. Thus one would recover the standard
picture. Clearly this is not the case of the brane models. What else
must be specified and why? This is an interesting question that we
hope to explore further. Our guess is that the naked singularities
introduce additional information that has been implicitly specified. In the
absence of a new general principle that specifies these additional
data one would have to give up the notion of causality (at least on a
global scale). This may be the price one must pay in order to solve
the cosmological constant problem.

\bigskip

{\it Acknowledgments} 
We would like to thank Csaba Cs\'{a}ki and Ira Rothstein for discussions.
This work is supported by the Department of
Energy under contract No.\ DOE-FG03-97ER40546.


\end{document}